\documentclass[twocolumn, amsmath,amssymb, superscriptaddress,nofootinbib]{revtex4-1}

\usepackage{dcolumn}
\usepackage{url}
\usepackage{bm}
\usepackage{amsmath}
\usepackage{amsfonts}
\usepackage{graphics}
\makeatletter
\g@addto@macro\appendix{\setcounter{figure}{0}}
\makeatother
\usepackage[pdftex]{graphicx}
\usepackage{times}
\usepackage{setspace}
\usepackage{verbatim}
\usepackage{color}   
\usepackage[normalem]{ulem}

\DeclareMathOperator*{\E}{\mathbb{E}}
\newcommand{\Var}{\mathrm{Var}}
\newcommand{\Cov}{\mathrm{Cov}}
\newcommand{\Cor}{\mathrm{Cor}}

\begin{document}

\title{On the Predictability of Future Impact in Science}
\author{Orion Penner$^{\dagger}$}
\affiliation{Laboratory of Innovation Management and Economics, IMT Institute for Advanced Studies Lucca,  55100 Lucca, Italy}
\author{Raj Kumar Pan$^{\dagger}$}
\affiliation{Department of Biomedical Engineering and Computational Science, Aalto University School of Science, P.O.  Box 12200, FI-00076, Finland} 
\author{Alexander M. Petersen$^{*}$}
\affiliation{Laboratory for the Analysis of Complex Economic Systems, IMT  Institute for Advanced Studies  Lucca,  55100 Lucca, Italy}
\author{Kimmo Kaski}
\affiliation{Department of Biomedical Engineering and Computational Science, Aalto University School of Science, P.O.  Box 12200, FI-00076, Finland} 
\author{Santo Fortunato$^{*}$}
\affiliation{Department of Biomedical Engineering and Computational Science, Aalto University School of Science, P.O.  Box 12200, FI-00076, Finland} 
\date{\today}

\begin{abstract}
Correctly assessing a scientist's past research impact and potential for future impact is key in recruitment decisions and other evaluation processes. While a candidate's future impact is the main concern for these decisions, most measures only quantify the impact of previous work. Recently, it has been argued that linear regression models are capable of predicting a scientist's future impact. By applying that future impact model to 762 careers drawn from three disciplines: physics, biology, and mathematics, we identify a number of subtle, but critical, flaws in current models. Specifically, cumulative non-decreasing measures like the $h$-index contain intrinsic autocorrelation, resulting in significant overestimation of their ``predictive power''. Moreover, the predictive power of these models depend heavily upon scientists' career age, producing least accurate estimates for young researchers. Our results place in doubt the suitability of such models, and indicate further investigation is required before they can be used in recruiting decisions.
\end{abstract}

\maketitle

\footnotetext[1]{ $^{\dagger}$ Authors contributed equally. $^{*}$ Correspondence should be sent to: petersen.xander@gmail.com or santo.fortunato@aalto.fi}

\section{Introduction}
Science has evolved a merit driven career advancement process in which an individual is promoted through the various career stages on the strength of his or her past achievements and perceived potential for future achievement. Committees charged with the task of evaluating the past accomplishments and projecting the future success of applicants are at the core of these advancement decisions, whether they be fellowship, grants, tenure track hires, tenure {\it etc.} In this context, evaluation is rarely a straightforward matter, as recent case studies indicate  that  grant committee selection decisions  do not necessarily correlate with either the peer-review process or cumulative achievement measures~\cite{Besselaar_past_2009}. 

Faced with applicant pools ranging in size from dozens, for tenure track hires, to thousands for national fellowship and tenure competitions, it is a great challenge to distill the contents of each {\it curriculum vitae} to an assessment of an individual's past, present and future impact and arrive to an appropriate ranking of candidates. Further, it is important to recognize that {\it future} impact is at the heart of this matter because the ultimate questions are: Which candidate will be most successful in {\it this} position? With {\it this} fellowship? Do the most with {\it this} grant? Emphasis is typically placed on past success but, for the most part, it is only relevant in so far as it correlates with future success.


When an early career scientist is selected for a tenure track position it is not simply a matter of filling an open position. The hire itself is an investment, at some institutions with low tenure rates it can amount to an outright bet on one researcher who requires a start-up package upwards of a millions of dollars \cite{stephan_how_2012}. The economics alone make this an issue that deserves attention. Nevertheless, beyond finances, these career advancement decision also play a critical role in most of the major problems commonly identified with the academic profession. For example, while gender biases may appear as early as undergraduate studies~\cite{moss-racusin_science_2012}, it is widely felt that 'pipeline' really leaks in the later career decision points~\cite{ginther_women_2004,beyond_2007,duch_possible_2012}.

\begin{figure}
\centering{\includegraphics[width=0.8\columnwidth]{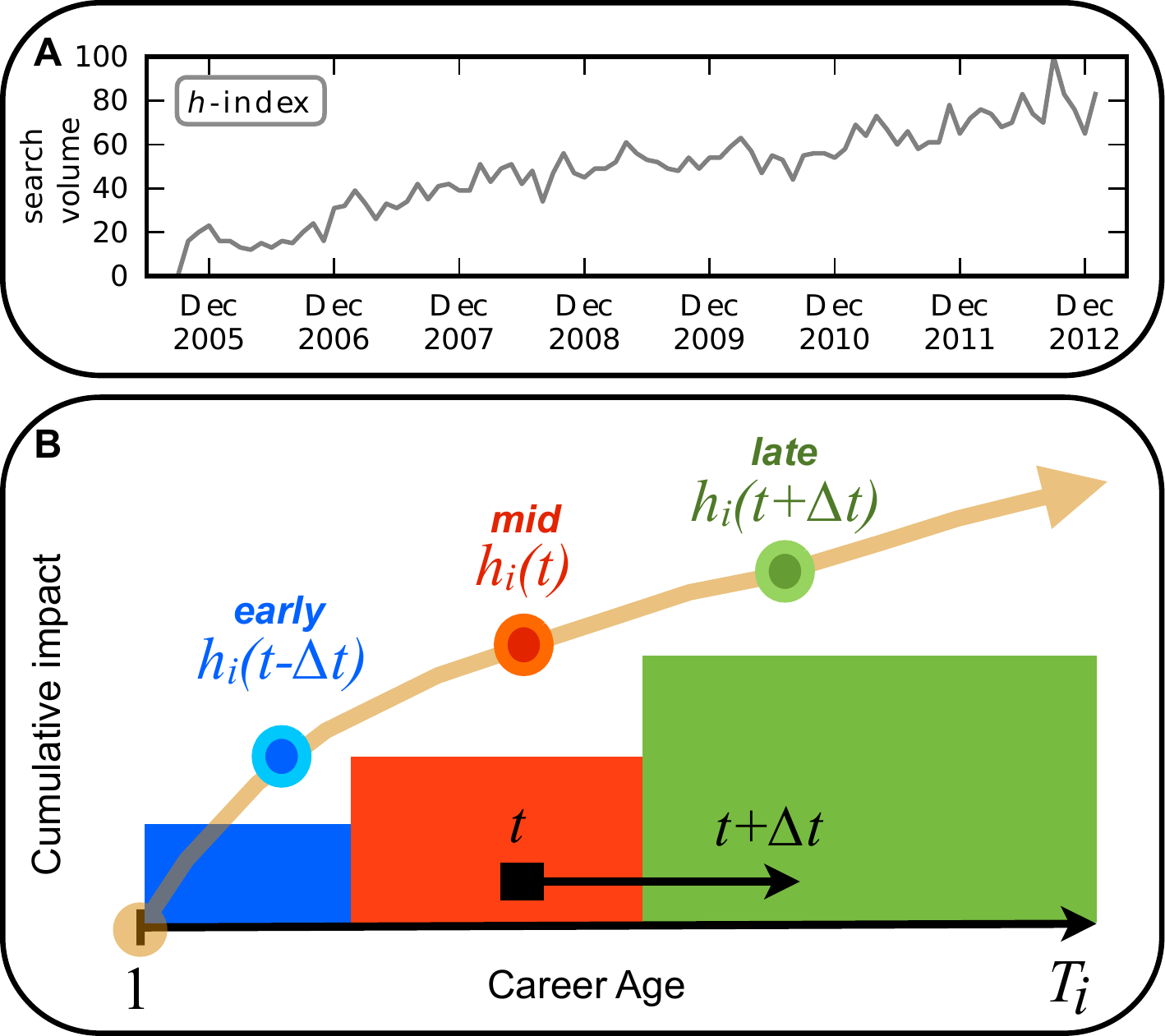}}
\caption{\label{fig:f1} (A) Monthly {\it Google} search volume for the term ``$h$-index'', normalized to \% peak value. Since the initial publication proposing the $h$-index on Nov. 15, 2005 \cite{hirsch_index_2005}, there has been roughly a 4-fold increase in $h$-index search volume over the 7-year period Dec. 2005 - Dec. 2012, capturing the persistent increasing interest and use of $h$ over time~\cite{GoogleTrendH}.
(B) Schematic illustration of the career stages that define academic careers. The $h$-index is a cumulative non-decreasing quantity intended to measure both the productivity and impact of a scientist $i$ up to year $t$~\cite{hirsch_index_2005}. However, models for predicting  $h(t+\Delta t)$ must account for two important factors: (i) $h(t)$ is non-decreasing so that ``predictability'' measures for $h(t+\Delta t)$ can be artificially inflated, and (ii) variations in the ``risk'' profile and the ``production function'' of scientists across career stages must be accounted for in predictive models.}
\end{figure}

For individual researchers the most widely known measure of impact is Hirsch's $h$-index~\cite{hirsch_index_2005}. Debate continues over whether $h$-index is a good way to measure a researcher's quality, but as it is evident by its growth in popularity~[Fig.\ref{fig:f1}~(A)] it is reaching a level of acceptance and more importantly, a level of formal use~\cite{ANVUR}. While it has been shown that a correlation exists between a researcher's current and future $h$-index, $h$-index is clearly a measure of a researcher's past accomplishments~\cite{hirsch_does_2007}. In recent work Acuna et al. propose a model for a researcher's future $h$-index and thereby establish a clear and concrete framework for connecting a researcher's current CV to his or her future impact in research~\cite{acuna_future_2012}. On the conceptual level this aligns much better with the goal of most career advancement decisions, as they are largely focused on what a researcher {\it will} produce rather than what he or she {\it has} produced.

However, on a technical level cumulative achievement models, such as the Acuna model, suffer from methodological flaws mainly arising from the fact that the $h$-index is a non-stationary measure~\cite{penner_commentary:_2013,schreiber_how_2013}.
Here we show that any regression model aimed at ``predicting'' should avoid using cumulative, non-decreasing, career measures because the retention of past information intrinsic to such measures will yield artificially large coefficients of determination $R^{2}$. A second methodological flaw exists in that prediction models should not mix career data from different age cohorts because such models deal poorly with the radically different levels of uncertainty characteristic of the various stages of career trajectories~[Fig.\ref{fig:f1}~(B)]. Even efforts to predictively model academic careers by disentangling the past and future components of scientific achievement~\cite{mazloumian_predicting_2012}, suffers from this second methodological flaw. 

Analyzing a large set of careers distributed across 3 disciplines, physics, biology and mathematics (see Methods), we show that although future measures of impact are correlated with past measures, the current state of the art models simply do not do a good enough job of predicting future impact to be used with confidence in the career advancement decision process. We demonstrate this using career data of established scientists, as well as junior scientists. The analysis of the benchmark set of stellar senior scientists serves as an upper bound on ``predictive power'', while the junior scientists represent a set closer to the typical case in which these will be applied in real academic hiring decisions.

\begin{figure}[t]
  \centering{\includegraphics[width=\columnwidth]{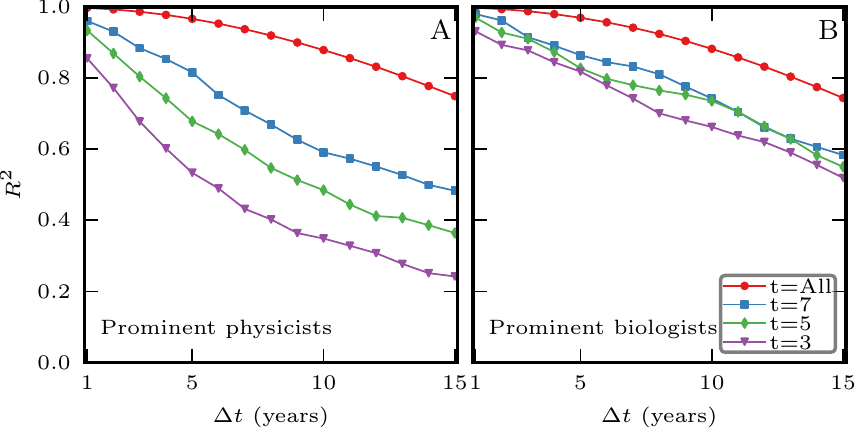}}
  \caption{The ``predictive power'' of the regression model of the $h$-index for different disciplines and for different career age cohorts (years since first publication $t = 3, 5, 7$). The curve for $t=$All shows the model of Eq.~\ref{eq:Acuna_model}, where all career ages were lumped together. For all the cases, overall regression model is significant ($p<10^{-6}$, calculated from F-statistic).}
\label{fig:Rsq_h}
\end{figure}
\section{Results}
\subsection{Modeling cumulative measures}
Here we consider linear regression models of the $h$-index but the analysis presented can be trivially extended to any cumulative measure of impact. A recent publication proposes a model for predicting an individual's future $h$-index based on linear regression of five other metrics~\cite{acuna_future_2012}. As a group, these five metrics were found to be the best for predicting future $h$-index.  In this linear regression model the $h$-index $h(t+\Delta t)$ of an individual at time $t+ \Delta t$ is given by
\begin{equation} \label{eq:Acuna_model}
  \begin{split}
h(t+\Delta t) &= \beta_{0}(\Delta t) + \beta_{h}(\Delta t) h(t) + \beta_{\sqrt{n_{\mathrm{p}}}}(\Delta t) \sqrt{n_{\mathrm{p}}(t)} \\
&+ \beta_{t}(\Delta t) t + \beta_{j}(\Delta t) j(t) + \beta_{q}(\Delta t) q(t) \ .
  \end{split}
\end{equation}
The variables found on the right-hand side of Eq.~\ref{eq:Acuna_model} are values calculated for a given $t$,  the number of years since the researcher's first publication. We will also refer to $t$ as ``career age''. For a given researcher, at a given career age $t$, the other variables are as follows: $h(t)$ is the $h$-index; $n_{\mathrm{p}}(t)$ is number of publications authored or co-authored; $j(t)$ is the number of distinct journals of the publications; $q(t)$ is the number of papers published in high impact journals. The parameter associated with each independent variable is arrived at using linear regression with elastic net regularization (see Methods). We apply the above model to predict the future $h$-index (as measured by the percentage variance explained, given by the squared correlation coefficient $R^2$) for both prominent physicists and prominent biologists. For both data sets the model shows high $R^2$ when lumping together all career ages (red curves in~Fig.~\ref{fig:Rsq_h}). Even 15 years into the future the model yields $R^2$ values of 0.75 and 0.76, respectively. These results are consistent with previous analyses and give the impression that the model is quite good at predicting a scientist's future $h$-index. For both these datasets, the variations of standardized coefficient are shown in Appendix Fig.~S1. The coefficient related to the $h$-index at the time of prediction (career age $t$) is the largest; the coefficient for the number of article published is also quite high especially in the distant future. In contrast, coefficients for publishing in many distinct journals and top journals are relatively small.

\subsection{Age-dependent cumulative model}
To assess the suitability of prediction models for applications in the real world, we analyze the $t$-dependence of the above model. We use the same regression variables as in Eq.~\ref{eq:Acuna_model} but disaggregate the prediction problem into sets of fixed career age ($t$). 
By modeling each career age separately we analyze the robustness of the above model with respect to varying career age. In this case the predicted $h$-index $\Delta t$ years in the future, of a scientist who is at a career age $t$, is given by:
\begin{equation} \label{eq:t_model}
  \begin{split}
h(t+\Delta t) &= \beta_{0}(t,\Delta t) + \beta_{h}(t,\Delta t) h(t) + \beta_{\sqrt{n_{\mathrm{p}}}}(t,\Delta t) \sqrt{n_{\mathrm{p}}(t)} \\ 
&+ \beta_{j}(t,\Delta t) j(t) + \beta_{q}(t,\Delta t) q(t) .
  \end{split}
\end{equation}
Note that as the data is already segregated by career age, $t$ is not considered as an independent variable in this version of the model. In Figure~\ref{fig:Rsq_h} we also show the model's predictive power for different career ages, for prominent physicists and biologists. The model's predictive power for early career researchers is far lower than the previous model where all career ages were lumped together ($t=$All). Although these results indicate the future of scientists at early stages of their career is less predictable, the $R^2$ values are still quite high, particularly for biologists. Those who are at the 3rd and 5th year of their career have $R^2=0.63$ and $R^2=0.73$ respectively, 10 years into the future. These values are notably high and may give the impression that an individual researcher's career trajectory is easily predicted even from a very early point. However, in the following section we show that cumulative measures like the $h$-index contain an intrinsic auto-correlation that not only results in this career age difference in the predictive power, but more importantly, to a dangerous overestimation of the model's overall predictive power. Further, the variations of standardized coefficients as shown in Appendix Fig.~S2 for $t=3$ and $t=7$ are different compared to the $t=$All case. Although, the coefficient related to the $h$-index is still largest, the coefficient for the number of papers in high impact journals is comparable, especially for biologist career. The variation of the coefficient related to $h$-index also increases with time, which is in contrast to the observation when all career ages were lumped together ($t=$All). Moreover, different coefficients for different career age means that they can not be aggregated together for regression analysis. Further, when a given dataset is sliced into two different groups, both the $R^2$ values as well as the coefficients of the regression models were different~(Fig.S3-S4), suggesting another weakness of this analysis.

\subsection{Non-stationary time series}
An academic career is an endeavor influenced by many factors, and in that light the Acuna model takes a step in the right direction by integrating several different variables into a prediction. However, the $h$-index is a {\it cumulative} measure and hence, is non-stationary. This makes $h$-index the incorrect dependant variable to target for prediction.
In this context we are using the weak definition of stationarity, which requires the mean and variance of a generic stochastic process to be time independent and the auto-covariance between the variable at $t$ and $t+\Delta t$ be a function only of $\Delta t$. As we show below its non-stationary nature makes $h$-index a poor predictor because it implies an intrinsic correlation that (i) explains, in part, the career age dependence noted above and (ii) results in an overestimation of the predictive power of models focused on predicting the future $h$-index and all other cumulative measures.

First, we consider a simple model for the evolution of an individual researcher's $h$-index, in which his/her $h$-index in a given year is a sum of yearly independent and random increments $\Delta h$. 
Hence, for a given researcher $s$, his/her $h$-index after $t$-years is given by
\begin{equation}
  h^{s}(t)=\sum_{i=1}^{t}\Delta h^{s}_i
  \label{eq:randomStepModel}
\end{equation}
where the $\Delta h^{s}_i$ are independent displacements with $\E (\Delta h^{s}_i ) = \mu_s$ and $\Var (\Delta h^{s}_i) = \sigma_s^2$, for all $i$. 

Next we consider the statistical properties of the above model. The expected value of the $h$-index at career age $t$ is 
\begin{equation}
  \E [h^{s}(t)] = \E \left[  \sum_{i=1}^{t}\Delta h^{s}_i \right]
  = \sum_{i=1}^{t} \E  [\Delta h^{s}_i]  = t\mu_s \ ,
  \label{eq:modelMean}
\end{equation}
and the variance
\begin{equation}
  \Var [h^{s}(t)] = \Var  \left[ \sum_{i=1}^{t}\Delta h^{s}_i \right]
  = \sum_{i=1}^{t} \Var [\Delta h^{s}_i] = t\sigma_s^2.
  \label{eq:modelVar}
\end{equation}
The auto-covariance is 
\begin{equation}
  \begin{split}
  \Cov [h^{s}(t+\Delta t),h^{s}(t)] &= \sum_{i=1}^{t+\Delta t}\sum_{j=1}^{t} \Cov(\Delta h^{s}_i,\Delta h^{s}_j)\\
  &= \sigma_s^2 \sum_{i=1}^{t+\Delta t}\sum_{j=1}^{t} \delta_{ij} = t\sigma_s^2.
  \end{split}
  \label{eq:modelCov}
\end{equation}
Thus, the correlation between $h(t+\Delta t)$ and $h(t)$ equals
\begin{equation}
  \begin{split}
  \Cor [h^{s}(t+\Delta t),h^{s}(t)] &= \frac{\Cov [h^{s}(t+\Delta t),h^{s}(t)]}{\sqrt{\Var{[h^{s}(t+\Delta t)]}\Var{[h^{s}(t)]}}} \\
  &= \sqrt{\frac{t}{t+\Delta t}}.
    \end{split}
  \label{eq:modelCor}
\end{equation}
The mean, variance and auto-covariance depend on $t$. Further, $h^{s}(t+\Delta t)$ and $h^{s}(t)$ are completely correlated when $\Delta t/t \approx 0$, that is when the researcher's career age is much greater than the number of years into the future you are attempting to predict his/her $h$-index. Likewise, $h^{s}(t+\Delta t)$ and $h^{s}(t)$ are completely uncorrelated as $\Delta t/t \to \infty$, i.e. when attempting to predict an individual's $h$-index many more years into their future than the current career age. 

Even disregarding the limiting behavior, Eq.~\ref{eq:modelCor} shows why regression models that attempt to predict the future $h$-index cannot perform as well for `young' careers as for `old' ones. Further, the fact that the correlation between current and future $h$-index intrinsically increases with researcher's age (for fixed $\Delta t$) indicates that the observed predictive power of models of $h(t+ \Delta t)$ may only be an outcome of general properties of the evolution of cumulative measures, rather than true ability to predict the future impact of a researcher. 

\begin{figure}
\centering{\includegraphics[width=\columnwidth]{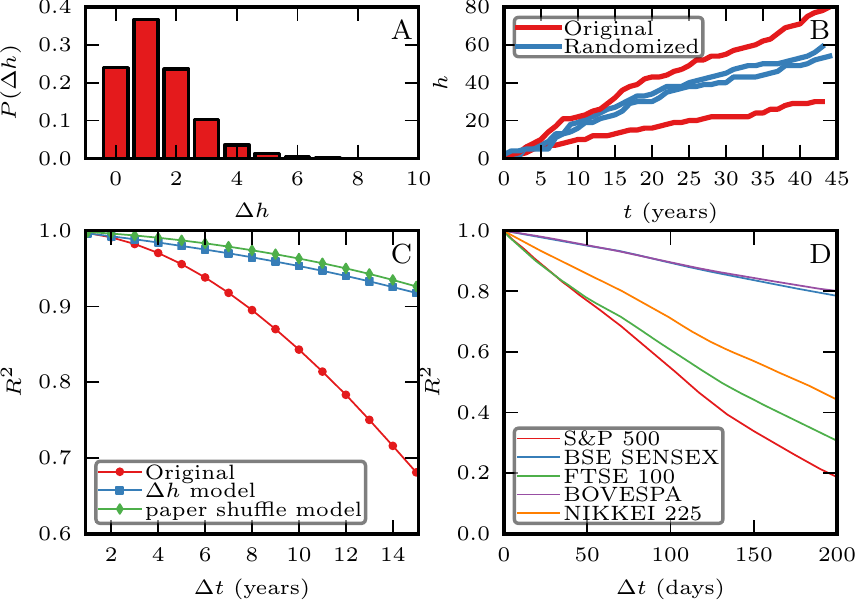}}
\caption{Correlation in non-stationary time series. (a) Distribution of $\Delta h$, i.e., the increment in scientist's $h$-index in consecutive years. (b) The evolution of $h$-index of two scientists in our dataset and their randomized version. (c) Variation of ``predictability'' $R^2$ with time for two different null models considered in the paper. (d) The auto-correlation of the actual value of the stock market index (not the price return) of 5 different countries. In (c) and (d), overall regression model is significant ($p<10^{-6}$).}
\label{fig:nullModels}
\end{figure}

\subsection{Empirical evidence of overestimation}
In this section we provide additional evidence that a trivial correlation is indeed present in $h$-index and it leads to a significant overestimation of the predictive power of linear models. To do this we resort to null models. That is, we explore a number of methods for constructing synthetic careers from the real career data, and show that when linear models for $h$-index are applied to these careers high $R^2$ values result. However, within these models all information that a linear regression model should be using to predict an individual's future $h$-index has been `scrambled', thus the resulting $R^2$ values should be (essentially) nil in the absence of the correlation arising from the fact $h$-index is a cumulative measure.

We refer to our first null model as the $\Delta h$ {\it null model}. Here we construct synthetic careers of physicists with the following procedure. First we generate the distribution of single year $h$-index increases for all careers in a given dataset. Figure~\ref{fig:nullModels}~(a) shows this distribution is narrow, with 98\% of the yearly increments less than 5. Second we generate a career by constructing a sequence of yearly $h$-index increases, drawn randomly from the distribution generated in the previous step. Two such career trajectories can be found in Figure~\ref{fig:nullModels}~(b). Finally we apply a simple linear model, $h(t+ \Delta t) = \beta_0 + \beta_h h(t)$. The $R^2$ values produced by this approach can be found in Figure~\ref{fig:nullModels}~(c). The $R^2$ values are quite high, far higher than the cumulative model of Eq.~\ref{eq:Acuna_model} applied to real careers. But what do these high $R^2$ values mean? Are they an indicator of predictive power and ability to discriminate between promising and not so promising careers? This is not the case as due to the manner in which the careers are generated, over any interval, the $h$-index of a researcher will increase by the same (average) amount at each step, regardless of whether the researcher has a high or a low $h$-index at that point. We conclude that such high $R^2$ values do not indicate predictive power, but they are rather evidence of intrinsic autocorrelation.

We refer to our second null model as the {\it paper shuffle null model}. In this case all papers published in year $t$ are shuffled and distributed randomly across all researchers (see Appendix for details). Hence, in this model the number of papers a researcher published in each year of his/her career is conserved. However, since papers are randomly assigned to each researcher each career is, statistically speaking, indistinguishable from each other in that every one has the same probability of `writing' a high impact paper. In Figure~\ref{fig:nullModels}~(c) it can be seen that, as with the $\delta h$ null model, this null model produces high $R^2$ values again indicating not predictive power but the presence of inherent correlation.

Finally, as an example of a system where simple models are known to have little predictive power, yet produce significant $R^2$ values, we turn to financial time series. We considered the stock market index of 5 different markets for the 15-year time period October 1997 to September 2012. In Figure~\ref{fig:nullModels}~(d) we plot the correlation (regression) of the index value at time $p(t+\Delta t)$ against $p(t)$ as a function of $\Delta t$. We note that this quantity exhibits a high degree of correlation even after 100 days. However, the analysis of the autocorrelation of index return (the actual predictability) shows that it decays quickly, thus supporting the efficient market hypothesis~\cite{malkiel_efficient_2003,mantegna_introduction_1999}.

\begin{figure*}[t]
\includegraphics[width=0.95\textwidth]{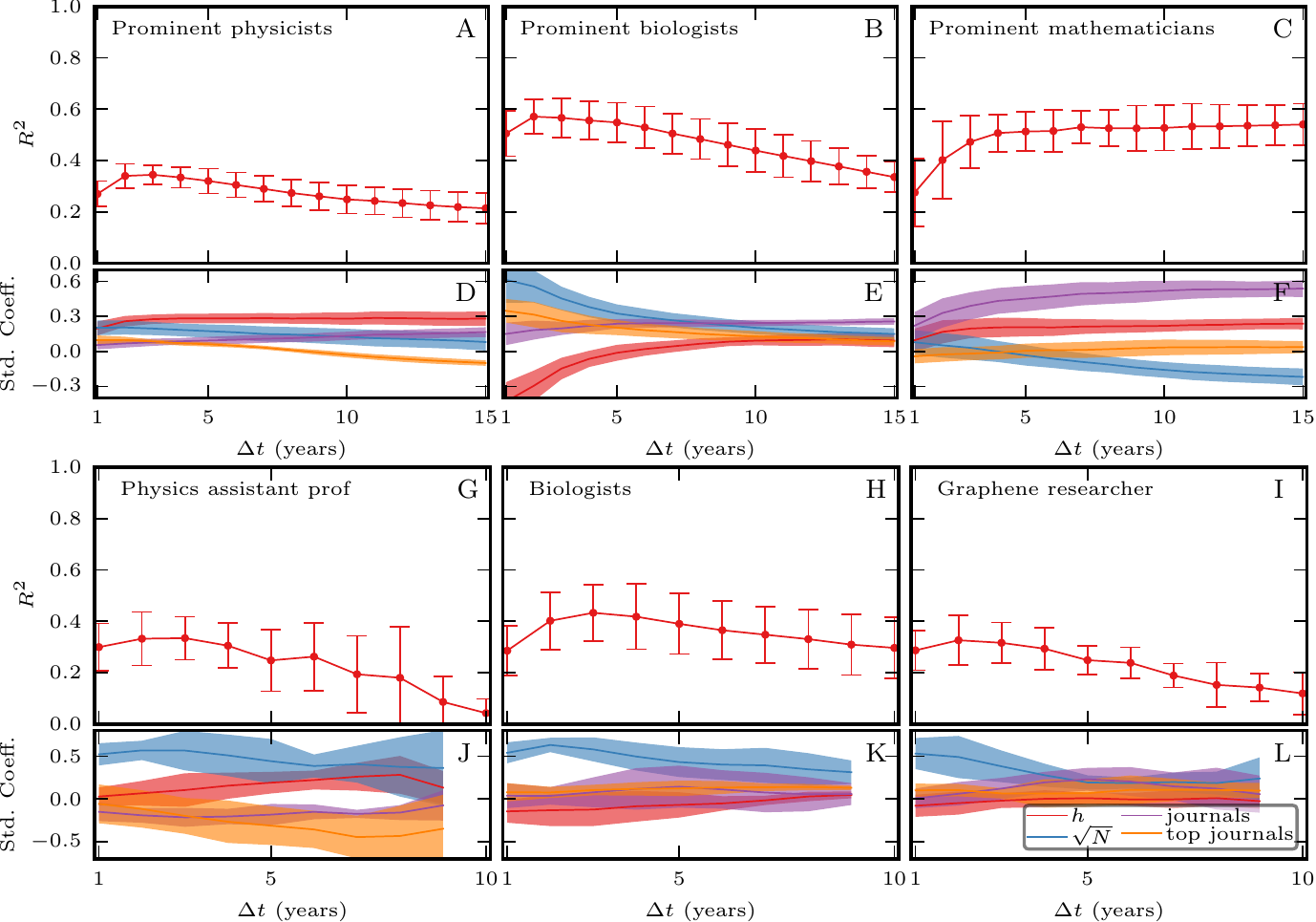}
\caption{The ``predictive power'' of $h$-index increments ($\Delta h(t,\Delta t)$) for different discipline. (A,B,C) Variation of the mean $R^2$ as a function of time period $\Delta t$ over which the increment is calculated for established physicists, biologists and mathematicians. The mean is calculated by averaging over different career age cohorts $t=2,\dots,15$. (D,E,F) Variation of the mean standard coefficient as a function of $\Delta t$. The shaded region indicates the 95\% confidence error bars. Similar plots are also shown for relatively young researchers in (G,H,I) for assistant professors in physics, biologists and graphene researchers. As the careers of young scientists are short, in this case the mean is calculated by averaging over different career age cohorts $t=2,\dots,8$. In all the cases, overall regression model is significant ($p<10^{-2}$).
}
\label{fig:Rsq_dh}
\end{figure*}
\subsection{Modeling non-cumulative measures}
The results presented above provide significant evidence that linear regression models are not so much predicting future impact as they are picking up on a correlation intrinsic to cumulative measures. Auto correlation, Eq.~\ref{eq:modelCor}, is only present in cumulative measures like total number of publications, total number of citations, total number of publications in distinct journals, {\it etc}. It is not present in non-cumulative measures, e.g.,  the incremental $h$-index, $\Delta h(t,\Delta t)= h(t+\Delta t) - h(t)$. Following the derivation above, the mean $\E [\Delta h(t,\Delta t)] = \mu\Delta t$ and variance $\Var [\Delta h(t,\Delta t)] = \sigma^2\Delta t$ are independent of time, resulting in the auto-covariance $\Cov [\Delta h(t+\tau,\Delta t),\Delta h(t,\Delta t)] = 0$ if  $\tau>0$. 
Hence, it is important to examine the $R^2$ for non-cumulative measures. Here we focus on a regression model for the incremental $h$-index $\Delta h(t,\Delta t)$ of a scientist at career age $t$, which by analogy with Eq.~\ref{eq:t_model} reads
\begin{equation} \label{eq:dt_model}
  \begin{split}
\Delta h(t, \Delta t) &= \alpha_{0}(t,\Delta t) + \alpha_{h}(t,\Delta t) h(t) + \alpha_{\sqrt{n_{\mathrm{p}}}}(t,\Delta t) \sqrt{n_{\mathrm{p}}(t)} \\ 
&+ \alpha_{j}(t,\Delta t) j(t) + \alpha_{q}(t,\Delta t) q(t) .
  \end{split}
\end{equation}
In Appendix Fig.~S5 we show this model's ``predictive power'', as measured by $R^2$, for different career ages $t$ and varying horizons $\Delta t$. All the curves except for early career years $t=1$ and $t=2$ follow similar behavior and there is no consistent trend of decreasing $R^2$ with decreasing $t$. The careers at $t=1$ show lower correlation, indicating that the state of an individual's CV after his/her first year of publishing is a poor predictor of his/her future trajectory. In Figure~\ref{fig:Rsq_dh} we show this average predictive power for the model when applied to established physicists, biologists and mathematicians at different age cohorts. It is immediately clear that when dealing with the non-cumulative measure, $\Delta h(t,\Delta t)$, the model has significantly less predictive power.

Figure~\ref{fig:Rsq_dh} also shows that the incremental variation in the $h$-index of a prominent biologist is more tightly connected to his/her past metrics. We speculate this may be due to other factors, like leading a large laboratory. We note similar behavior for prominent mathematicians. As these three datasets represent only prominent scientists, selected based upon their high success, the $R^2$ values give an upper bound on predictability of scientists in that field. In contrast the dataset of physics assistant professors, young biologists and graphene researchers, all relatively young scientists, exhibit much lower $R^2$. Finally we show the variation of the mean of the standard coefficient of the model. The coefficient related to $h$-index is not as important as we found for Eq.~\ref{eq:Acuna_model}, and other factors such as number of publications, number of publications in distinct journals, and number of publications in top journals are more important. For prominent biologists the coefficients for publication in top journals and number of publications are higher than for physicists. For mathematicians the coefficient related to the number of distinct journals is largest. In relative terms, the coefficient of the $h$-index is more important for physicists. 

Although this figure shows the average trend, one ought to exercise caution in interpreting the results because coefficients for scientists at different stages of their careers are also different. For example, Appendix Fig.S6 shows the coefficient for age $t=3$, $t=5$ and $t=10$ for both prominent physicists and biologists. It is easy to see that the coefficient related to the number of papers decreases as $\Delta h$ is measured over larger $\Delta t$. Further, for biologists, the coefficient for the number of publications in top journals is larger in the late part of the career than in the early stages. Nevertheless, the coefficients of the regression analysis were different even when for the same set of scientist during different age of their career. This variation in the coefficients across fields, as well as across career stages, indicates that it is unlikely there is a unique set of parameter that can be used to predict future impact for all cases.

\begin{figure}
\centering
\includegraphics[width=\columnwidth]{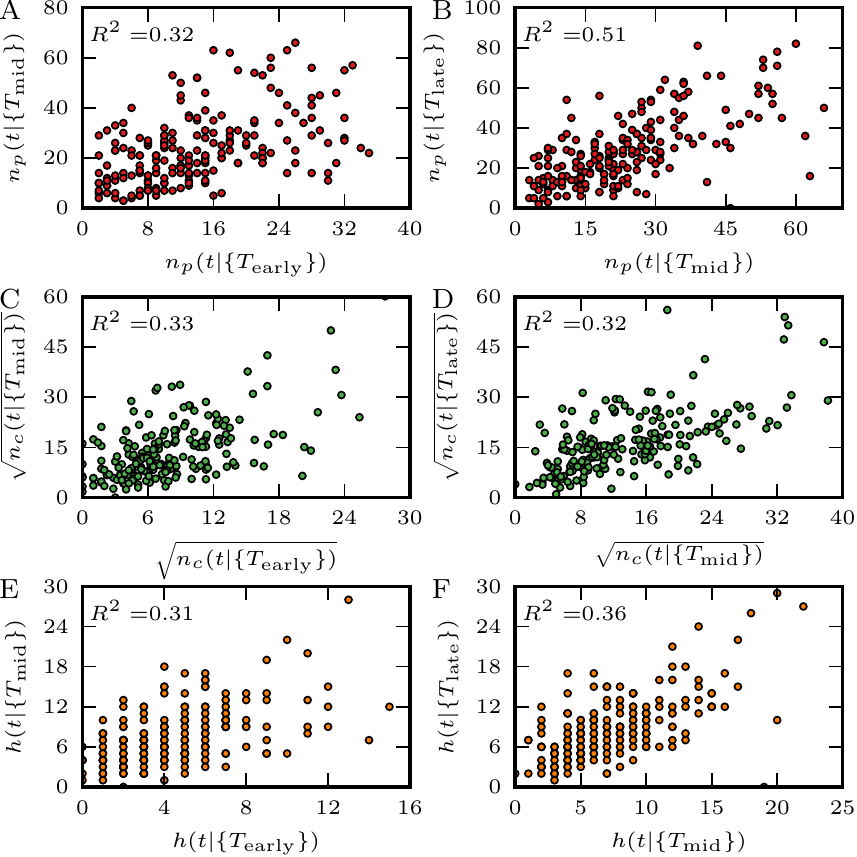}
\caption{Correlation between the past and the true future for prominent physics careers. Scatter plot of the number of papers calculated for each author using non-intersecting sets of papers published in (A) ``early''- and ``mid''- career periods and (B)``mid''- and ``late''- career periods. Scatter plot of the square root of number of total citations for each author in (C) ``early''- and ``mid''- career periods and (D)``mid''- and ``late''- career periods. Scatter plot of non-cumulative $h$-index $h(t \vert \{T_{j}\} )$ calculated for each author using non-intersecting sets of papers published in consecutive (E) ``early''- and ``mid''- career periods and (F)``mid''- and ``late''- career periods. The correlation coefficient $R$ for each plot is also shown.
}
\label{fig:trueFuture}
\end{figure}

\subsection{Correlating past and true future}
Although in the previous section we considered non-cumulative measures of scientific productivity and impact, the correlation between an individual's past accomplishments and future achievements deserves a more fine grained examination. For example, the number of citations received by a scientist at career age $t$, during the period $\Delta t$ years into the future depends both upon the papers he/she has written up to year $t$ and upon the papers published up to year $t+\Delta t$. Similarly, the increase in $h$-index during any given period is due to citations to papers he/she has already written in past years as well as citations to papers published during the period in question. In order to investigate the career uncertainty across academic transition points we analyze each scientist's citation impact over 3 consecutive non-overlapping periods. The first period, $\{T_{\mathrm{early}}\}$, starts at the beginning of his/her career, $t=1$, and extends up to $t=5$. The second period, $\{T_{\mathrm{mid}}\}$, starts at year $t=6$ and extends to $t=10$, while the third period, $\{T_{\mathrm{late}}\}$, starts at year $t=11$ and extends to $t=15$ years. For each period, we collect for each scientist only the publications that he/she published within that period, and, considered the citations received by these publications within the same period.

We calculate three non-cumulative impact measures for each scientist: (a) the total number of publications $n_{\mathrm{p}}(t \vert \{T_{j}\} )$; (b) the square root of total number of citations $\sqrt{n_{\mathrm{c}}(t \vert \{T_{j}\} )}$; (c) the $h$-index $h(t \vert \{T_{j}\} )$. These measures account only for citations within the period to papers also published within that period. In this way, we test the predictability of the citation impact of a scientist's  future work using publication information measuring his/her earlier research. Figure~\ref{fig:trueFuture} shows a scatter plot of physicists for all the three measures. The left panels show the correlation between the `early' and the `mid' career and the right panels show the correlation between the `mid' and the `late' career. The correlation coefficient $R$ is also shown for each measure. These values are lower than, but qualitatively similar to, the observation in Fig.~\ref{fig:Rsq_dh}, indicating that future measures are indeed somewhat correlated with the past. We found that for all the measures the correlation between past and future is similar. Thus our analysis suggests that all these measures are equally good (or equally bad) in predicting future impact. Further, the correlation between mid and late career is slightly higher than the correlation between early and mid. This is reasonable in so far as there is greater fluctuation in the early career stage than the later stages when scientists are more established. Additionally, our results diverge from recent work showing that future citations to future work are hardly predictable~\cite{mazloumian_predicting_2012}. Instead, we found low but significant correlation between past and future measures. It is possible that this difference arises from the fact that this portion of our analysis focuses on scientists that are all relatively well established, thus missing scientists that produce low impact work and ultimately exit academia. This result does nevertheless suggest that the predictability of top scientists can be used as an extreme upper bound for the predictability of all scientific careers. The results for prominent biologists and mathematicians are qualitatively similar, whereas for young researchers, physics assistant professors, young biologists and graphene researchers correlation is much smaller (Fig.S7-S11). 

\section{Discussion}
The sheer amount of information that enters into an evaluation is daunting. In addition to the research output, factors such as the prestige of an applicant's previous institutions  \cite{long_productivity_1978,long_entrance_1979}, supervisors \cite{amaral_mentor_2007}, volume and quality of service work, teaching and mentoring potential, {\it etc.}, are also  considered in the process. Indeed, science is based upon systems of reputation, which is typically estimated using cumulative measures~\cite{petersen_reputation_2013}. However, evaluation criteria that are heavily weighted on cumulative achievement measures  may reinforce  stratification and cumulative advantage mechanisms in science  \cite{cole_social_1973,hargens_structural_1984,jones_multi-university_2008,petersen_quantitative_2010}, which may inadvertently increase the risk burden of young careers \cite{petersen_persistence_2012}.


Thus, we need to not only understand the success and attrition rates of scientific careers, but, it is critical to grasp the limits-of-prediction. In the past, research, and especially researchers, have been evaluated qualitatively but now quantitative approaches, based upon citation counts, are becoming increasingly common. Indeed they are now being used formally and informally in the career advancement process. Citation counts, like other science metrics, are just one of the many dimensions of academic success and have to be used together with, and not instead of other evaluations. Still, if one wants to use science metrics in real comparative career evaluations, it is necessary to account for their biases and possibly correct them \cite{radicchi_universality_2008}.

Our analysis shows  that for the purpose of predicting a scientist's future $h$-index linear regression models suffer a variety of flaws. Their performance strongly depends upon career age. Cumulative, nondecreasing, dependent variables contain an intrinsic correlation that makes $R^2$ a misleading measure of predictive power. Removing this correlation by reformulating the problem as one of predicting the $h$-index increase ($\Delta h$) over a fixed time interval, and segregating the careers into different age cohorts, linear models do a poor job of predicting future impact. Finally, our effort to examine the correlation between the impact of a scientist's past papers and future papers shows there may be a relationship to be discovered, but doing so will require a highly sensitive and powerful approach.

Despite these shortcomings, and in fairness to those that have broken this path, the real impact of these early models does not, necessarily, lie in their ability to predict future impact. A significant contribution has been made by turning the critical eye of the community on the issue of predicting future scientific impact, and a much larger set of issues surrounding the use of quantitative measures in the academic career advancement process. But much work remains to be done before predictive models of future impact come of age and there are several obvious directions for future inquiry. For example, how the weights of coefficients vary, across disciplines as well as career ages should be thoroughly studied. 
As well, other independent variables should be studied in detail, for example, what impact does advisor prestige have upon a scientist's future $h$-index.

Of course, critical to all future efforts is the availability of high quality career data and some new, interesting, opportunities lie in that direction~\cite{ORCID}. The questions that could drive the future research are: What would the perfect prediction model need to be capable of in order to be suitable for real world application? Further, what additional characteristics would it need to have to see widespread and responsible use? 

With regards to the first question, it is critical that efforts to model future research impact focus on the fact that we are not predicting an individual's future impact in a vacuum. The vast majority of 'real' world uses demand models be able to differentiate between researchers, to correctly rank them in order of their future impact. The capacity to produce a correct ranking, not just a number for each researcher, is really what is critical. Indeed it is advisable that future work on predicting future impact bypass $R^2$ all together in favor of ranking based measures of predictive power. Turning to the second question, it is important that these rankings must be highly precise and explicitly assign a confidence score to the order. It is also highly desirable that these models be easy to calibrate because, as shown above, it is not possible for a single set of parameters to transcend the wide range of citation, publishing, {\it etc.} behaviors known to exist between disciplines. Hence, ease of calibration would be particularly important for adoption in less quantitative disciplines. It is also important that the community develops models that are able to separately predict future impact arising from future citations to past papers, and future impact arising from future citations to future papers. This may seem a minor distinction, but it is really at the heart of many hires, a tenure track position being a good example. In that case a candidate whose $h$-index will increase due to work performed in the position is far more desirable than one whose $h$-index increases due to work performed previously, assuming they both end up with the same $h$-index.

In closing, cumulative measures of future impact are not appropriate targets of predictive modeling because they contain trivial correlation by construction. We have provided significant evidence that the current predictive models for future impact possess far less predictive power than previously reported. Further, the next generation of efforts to predictively model future impact need aimed more directly at applications in the career advancement decision process.

\section{Methods}
\subsection{Data description}
We analyzed the publication profiles for  762 scientists divided into 3 broad disciplines: 476 physicists, 236 cell biologists, and 50 pure mathematicians. The top-cited scientists in their respective field comprise the ``prominent'' scientist datasets. For each scientist we compiled his/her comprehensive publication and citation profile using the Thomson Reuters (formerly ISI) Web of Knowledge historical publication and conference proceedings database. For more information on author selection and disambiguation method, see the Appendix.

We also studied five different stock market indices each from a different country (a) S\&P 500 from US (b) BSE Sensex from India (c) FTSE 100 from UK (d) BOVESPA from Brazil and (e) NIKKEI 225 from Japan. The data was downloaded from {\tt www.finance.yahoo.com} and covers the period from  October 1997 to September 2012.

\subsection{Elastic net regularization for linear regression}
When the independent variables of a linear regression model are correlated the estimated coefficients obtained by least-square method are highly sensitive to random errors in the observed response. To resolve this problem we use elastic-net regularization which is useful when there are multiple features which are correlated with one another (collinear)~\cite{zou2005regularization}. There are two parameters, the first one is the mixing parameter $\alpha$, which controls the collinearity of the parameters, the second one is the regularization parameter $\lambda$, which controls the complexity of the model. In all our analysis we set $\alpha=0.2$, whereas the best $\lambda$ is determined by cross-validation. We also checked that our results are qualitatively similar for other alpha values, say $\alpha=0.1$. 
More information on all aspects of the regression method can be found in the Appendix.

\begin{acknowledgments}
\noindent  O.P. acknowledges funding from the Social Sciences and Humanities Research Council of Canada. Certain data included herein are derived from the Science Citation Index Expanded, Social Science Citation Index and Arts \& Humanities Citation Index, prepared by Thomson Reuters, Philadelphia, Pennsylvania, USA, Copyright Thomson Reuters, 2011.
\end{acknowledgments}

\bibliography{predictability}

\appendix

\section{Methods}
\subsection{Disambiguation strategy}
The ``disambiguation problem'' is a major hurdle in the analysis of careers, as multiple authors having the same initials, and even the same complete name, can appear as a single author. Here we use  disambiguated  ``distinct author'' data from Thomson Reuters Web of Knowledge, {\tt isiknowledge.com} using their matching algorithms to identify publication profiles of distinct authors. Further, we use its website portal {\tt ResearcherID.com}, where users upload and maintain their publication profiles. This ISI online database is host to comprehensive data that is well-suited for developing testable models for scientific impact \cite{petersen_methods_2010,petersen_statistical_2011} and career achievement \cite{petersen_quantitative_2010,petersen_persistence_2012}. 

\subsection{Selection of scientists}
We seek to compare variations in productivity and impact across distinct scientific fields as well as within fields. To do this we analyze a total 762 scientists  divided into 3 broad disciplines:  476 physicists,  236 biologists, and 50 mathematicians. 

{\bf Dataset A:} For the selection of high-impact physicists, we aggregate all authors who published in {\it Physical Review Letters} {\it (PRL)} over the 50-year period 1958-2008 into a common dataset. From this dataset, we rank the scientists using the citations shares metric defined in \cite{petersen_methods_2010}, and select the top 100. Such metric divides equally the total number of citations a paper receives among the $k$ coauthors, and also normalizes the total number of citations by a  time-dependent factor to account for citation variations across time and discipline. Hence, for each scientist in the {\it PRL} database, we calculate a cumulative number of citation shares received from only their {\it PRL} publications. This tally serves as a proxy for his/her scientific impact in all journals. We also choose from our ranked {\it PRL} list, randomly, 100 additional highly prolific physicists. The selection criteria for that dataset is that an author must have published between 10 and 50 papers in {\it PRL}~\cite{petersen_statistical_2011}. This likely ensures that the total publication history, in all journals, be on the order of 100 articles for each author selected. These two lists were curated in such a fashion in a previous publications, and as both of them represent prominent physicists, are merged and analyzed together for some of the analysis.  The average $h$-index of these 200 scientists is $\langle h \rangle = 56 \pm 20$.  

{\bf Dataset B:} For the selection of high-impact cell biologists we choose the top 100 careers based on publications in the journal {\it CELL}. These scientist's have average $h$-index of $\langle h \rangle = 97 \pm 34$. 

{\bf Dataset C:} For the selection of high-impact mathematicians we selected the 50 authors with the most publications in the prestigious journal {\it Annals of Mathematics}. The average $h$-index of these scientist is $\langle h \rangle = 20 \pm 10$. We choose only 50 since the variation in collaboration and productivity across mathematics is significantly smaller than in the experimental and theoretical natural sciences. 

The above three datasets consist of high-impact senior scientists with average academic age $40 \pm 11$, $40\pm 7$ and $63\pm23$, respectively.  

{\bf Dataset D:} We also consider 100 relatively young assistant professors from physics. To select the scientists in this dataset, we choose two assistant professors from each of the top 50 U.S. physics and astronomy departments ranked according to the magazine {\it U.S. News}. The average $h$-index of these scientists is $\langle h \rangle = 15 \pm 7$. 

For datasets [A]-[D] we used the ``Distinct Author Sets" function provided by ISI in order to increase the likelihood that only papers published by each given author are analyzed. On a case by case basis, we performed further author disambiguation for each author. Other datasets are comprised of a broad range of scientists with profiles on {\it ResearcherID.com} who satisfied the criterion of having more than 10 publications. 

{\bf Dataset E:} This dataset consists of 174 scientists who have published in the field of graphene research. Additionally, we also include 2 notable leaders of this field (Nobel Prize Laureates A. K. Geim and K. S. Novoselov). The average $h$-index of this group is $\langle h \rangle = 12 \pm 11$.

{\bf Dataset F:} This dataset consists of 60 ``molecular biology'' and 76 ``neuroscience'' ResearcherID scientists. We assume that such ResearcherID scientists have uploaded a representative (approximately update and complete) set of publications. The average $h$-index of the scientists in this group is $\langle h \rangle = 17 \pm 15$.

The last three datasets consist of relatively young scientists with average academic age $11 \pm 6$, $10\pm 7$ and $17\pm9$, respectively. In summary, we group the 762 scientists that we analyze into 6 sets.  We downloaded datasets A in Jan.  2010, B and C  in Apr. 2012, D  in Oct.  2010, and E, F in October 2012. 

\begin{figure}
\centering
\includegraphics[width=\columnwidth]{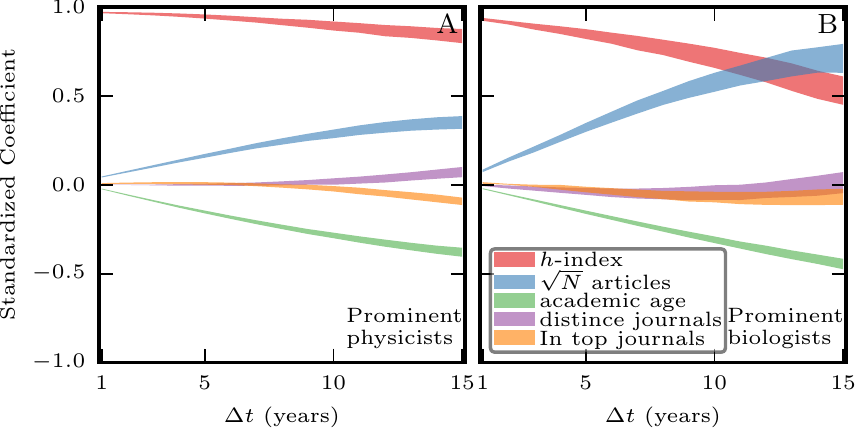}
\caption{Variation of the standard coefficient indicating the change in the contribution of each factor over time for different disciplines. The shaded region indicates the 95\% confidence error bars. The standard coefficients are shown for $t=$All case, where all career ages were lumped together.}
\label{fig:Rsq_h_CI}
\end{figure}

\begin{figure}[h!]
\centering
\includegraphics[width=\columnwidth]{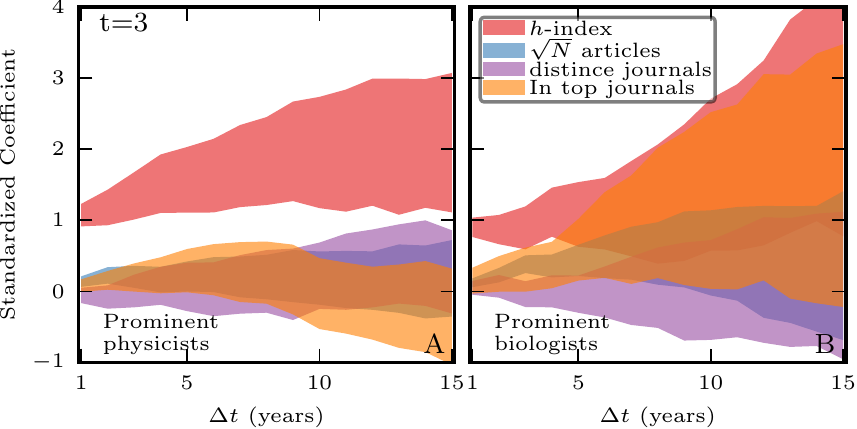}
\includegraphics[width=\columnwidth]{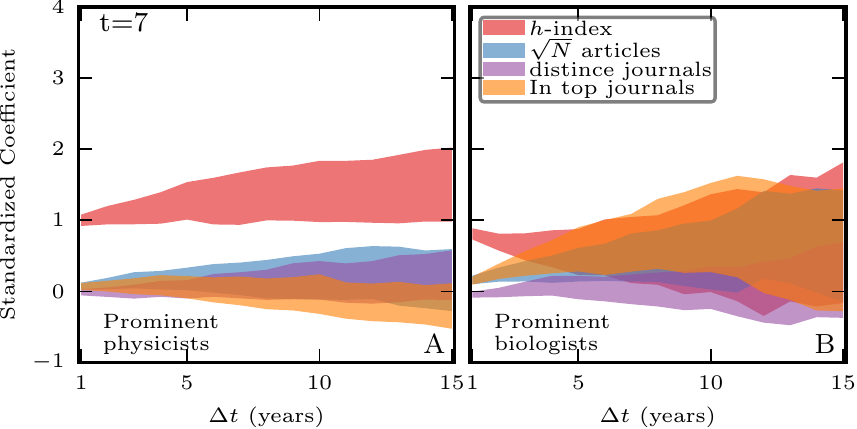}
\caption{Variation of the standard coefficient indicating the change in the contribution of each factor over time for different disciplines. The shaded region indicates the 95\% confidence error bars. The standard coefficients are shown for different age cohorts $t=3,7$.}
\label{fig:Rsq_h_CI}
\end{figure}

\begin{figure}
\centering
\includegraphics[width=\columnwidth]{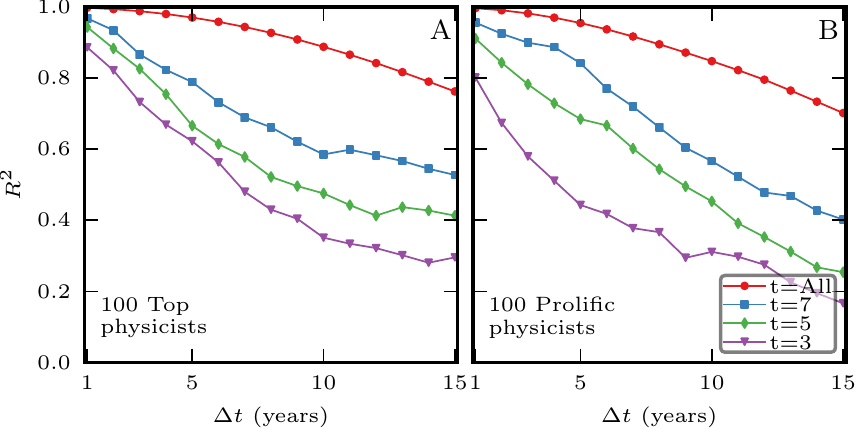}
\caption{The ``predictive power'' of the regression model of the $h$-index for physicist for different career age cohorts (years since first publication $t = 3, 5, 7$), with (A) 100 top and (B) 100 prolific physicists plotted separately. In the curve for $t=$All, all career ages were lumped together. For all the cases, overall regression model is significant ($p<10^{-6}$, calculated from F-statistic). The first group show slightly larger ``predictive power'' as compared to the second group, especially for large $\Delta t$.}
\label{fig:Rsq_h_SI}
\end{figure}

\section{Elastic Net regularization}
Regression analysis is a statistical technique for estimating the relationships among the dependent and independent variables. If $\mathbf{X}$ represent the independent variable and $y$ represent the dependent variable then the regression model relates these two variables as
\begin{equation}
  y=f(\mathbf{X},\mathbf{w}),
\end{equation}
where $\mathbf{w}$ are the unknown parameters. If the dependent variable is expected to be a linear combination of the independent variables, then in mathematical notation, the predicted value is expressed as
\begin{equation}
  \hat{y}=w_0+w_1x_1+\dots+w_px_p
  \label{eq:lRegress}
\end{equation}
Here, $w_0$ is the intercept and $(w_1,\dots,w_p)$ are the coefficients of the model. In general one can use an ordinary least square method to fit a linear model with coefficients to minimize the residual sum of squares between the observed responses in the dataset, and the responses predicted by the linear approximation. Thus, it can be represented as
\begin{equation}
  \min_w|| \mathbf{X}\mathbf{w}-y ||^2_2.
  \label{eq:leastSq}
\end{equation}
However, coefficient estimates for ordinary least squares rely on the independence of the model terms. When terms are correlated (also termed as collinear) this method becomes highly sensitive to random errors in the observed response, producing a large variance. Further if the number of features is large, it is possible to reduce the complexity of the model by forcing some coefficients to be small or zero. The elastic net regularization does this by imposing preferred solutions with fewer parameter values, effectively reducing the number of variables upon which the given solution is dependent. This method can be mathematically represented as 
\begin{equation}
  \min_w \frac{1}{2n}|| \mathbf{X}\mathbf{w}-y ||^2_2 +\lambda \alpha ||\mathbf{w}||_1 + \frac{\lambda(1-\alpha)}{2}||\mathbf{w}||^2_2
  \label{eq:elasticNet}
\end{equation}
Here, $||\dots||_1$ and $||\dots||_2$  are the $L^1$ and $L^2$ norm of the vector respectively. $\lambda \geq 0$ is a complexity parameter that controls the amount of shrinkage: the larger the value of $\lambda$, the greater the amount of shrinkage and thus the coefficients become more robust to collinearity. The parameter $\alpha$ controls how much collinearity is expected between features. In all our analysis we set $\alpha=0.2$, whereas the best $\lambda$ is determined by cross-validation. We also checked that our results are qualitatively similar for other alpha values, say $\alpha=0.1$. 

\section{Null models and randomized careers}
We used two different shuffling methods to create a set of randomized career profiles that do no have any inherent correlations. We use these careers as a benchmark for determining whether the predictability in the regression model is due to correlations in the scientific careers or to the career measure used.\\
{\it Paper shuffle model:} In this null model we start by shuffling papers only within a specific career age $t$. To be specific, let the paper repository $\{p\}_{t}$ be the set of papers $p$ published among all authors $i=1...A$ in career year $t$. 
To distribute papers to ``randomized'' career profiles, we randomly assign papers from the set $\{p\}_{t}$ until each author has $n_{i}(t)$ papers, where $n_{i}(t)$ is the value observed in his/her real career. This method approximately retains the collaboration patterns of each scientist (and his/her sub-discipline) which are largely responsible for growth in $n_{i}(t)$ over the career~\cite{petersen_persistence_2012}.
We tested this shuffling method by pooling together 200 prestigious physicists from datasets [A] and [B] into a single paper repository. 
Distinguishing papers according to year $t$ we approximately retain the properties of ``older'' papers versus ``younger'' papers, while ``washing out'' the historical, aging, and reputation effects that make empirical career profiles author specific. Using this repository we constructed 40,000 synthetic careers profiles used in our analysis.\\
{\it $\Delta h$ model:} In this null model we first obtain the distribution of single year $h$-index increases for all careers in a given dataset. Next, we generate a career by constructing a sequence of yearly $h$-index increases, drawn randomly from the distribution generated in the previous step. This null model is not based on the number of papers published by a specific scientist but rather on the length of career of each scientist, which is kept fixed as in the original dataset.

\begin{figure*}
\centering
\includegraphics[width=0.49\linewidth]{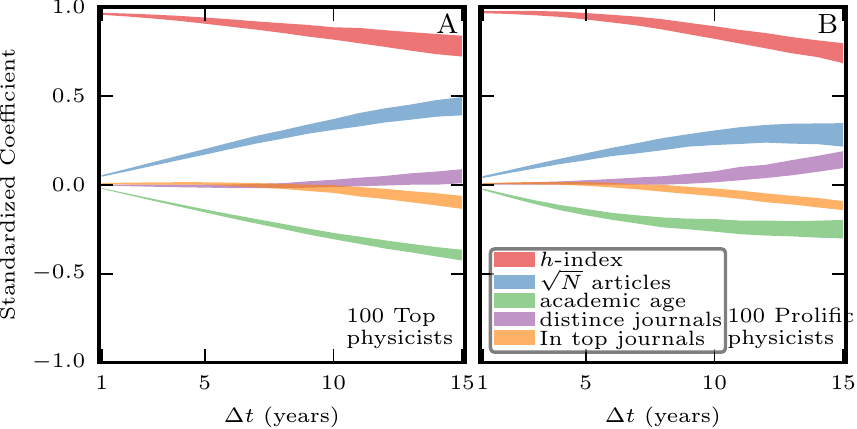}
\includegraphics[width=0.49\linewidth]{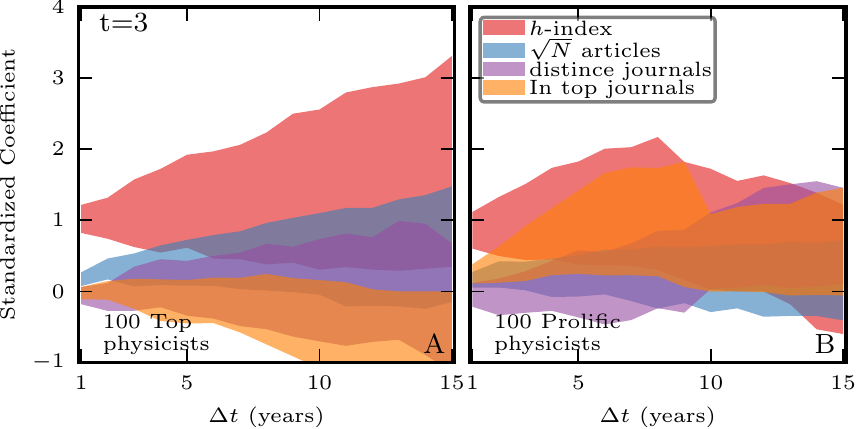}
\includegraphics[width=0.49\linewidth]{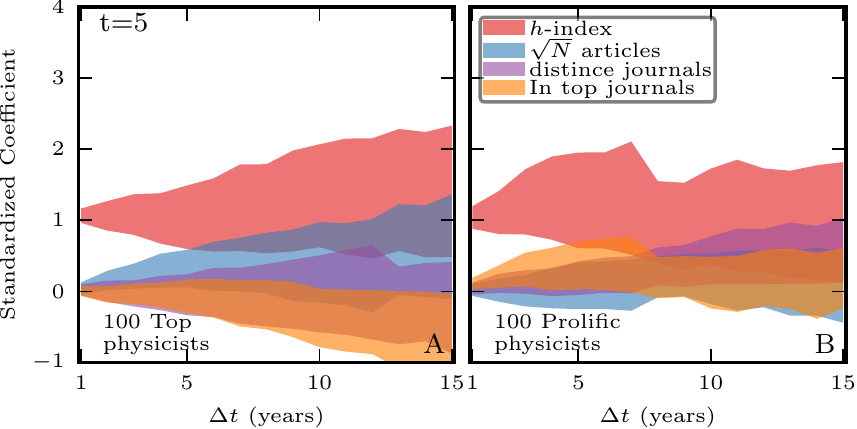}
\includegraphics[width=0.49\linewidth]{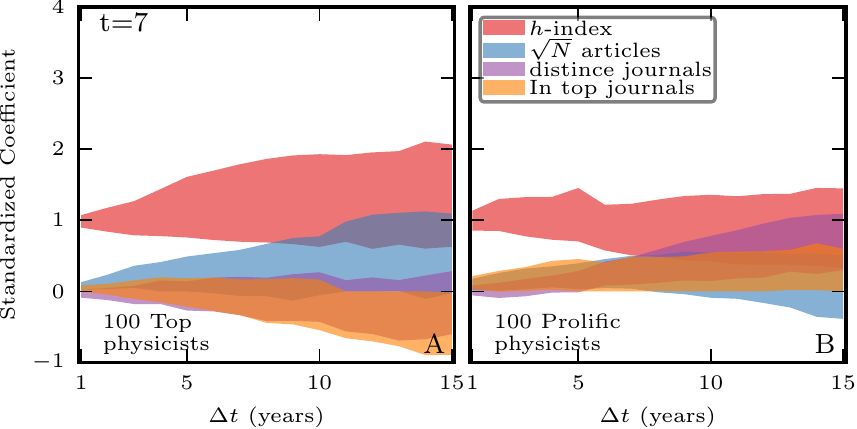}
\caption{Variation of the standard coefficient indicating the change in the contribution of each factor over time for physicist, with 100 top and (B) 100 prolific physicists modeled separately. The shaded region indicates the 95\% confidence error bars for different career age cohorts (years since first publication t = 3, 5, 7) for both the groups. The spread of the coefficients within a group and the differences in the coefficients across the two groups, indicate that parameters from one group is not suitable to predict the other group.}
\label{fig:Rsq_h_CI}
\end{figure*}

\begin{figure*}
\centering
\includegraphics[width=0.88\linewidth]{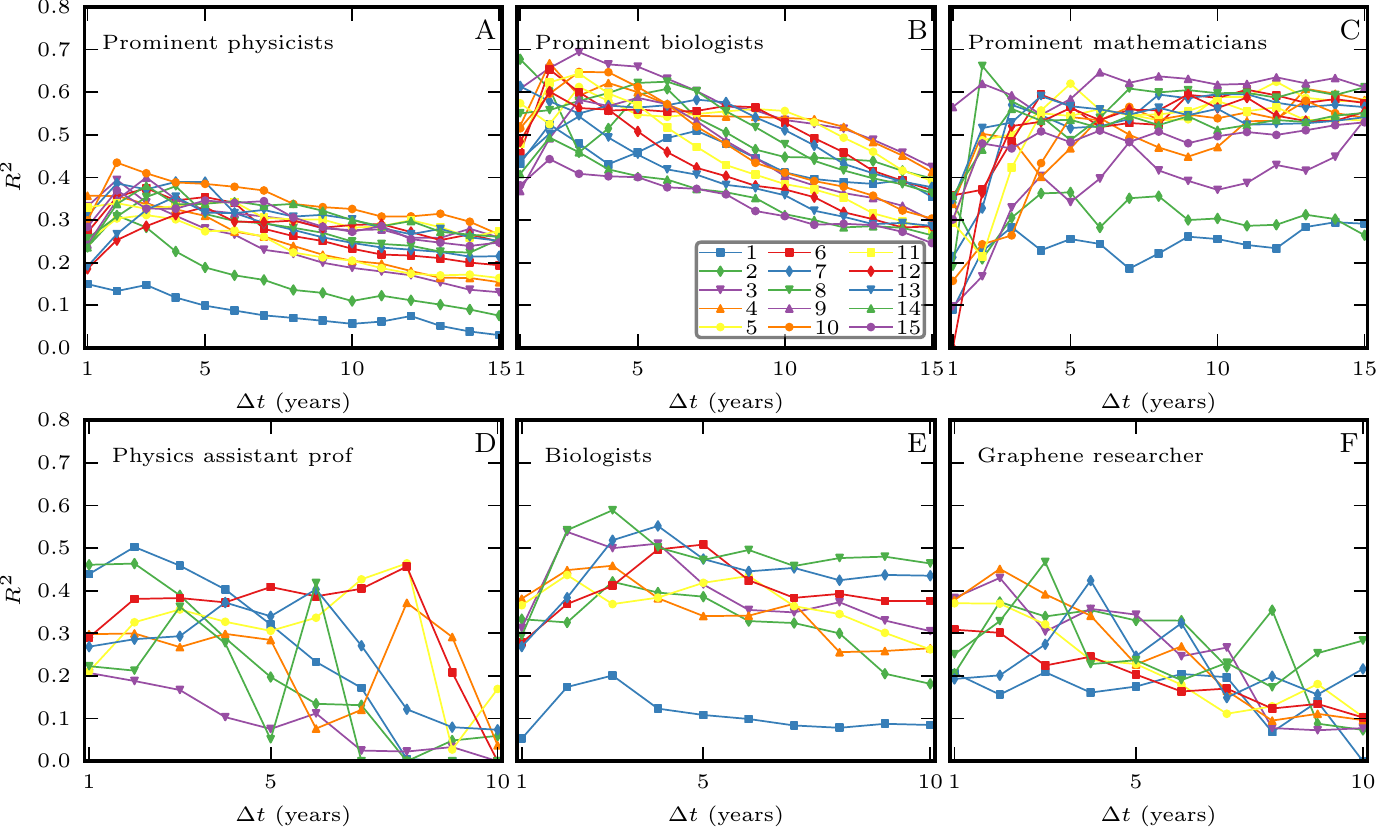}
\caption{The ``predictive power'' of $h$-index increments $\Delta h(t,\Delta t)$ for different disciplines and age-cohorts. For prominent physicists, biologists and mathematicians $t=1,\dots,15$. As the careers of young scientists are short, for the assistant professors in physics, biologists and graphene researchers
  $t=1,\dots,8$. In all the cases, overall regression model is significant ($p<10^{-2}$). For young researchers, as there were very few career with $t>8$, using regression model leads to over fitting and non-significant results ($p>.05$). 
}
\label{fig:Rsq_dh_age}
\end{figure*}

\begin{figure}[h!]
\centering
\includegraphics[width=\columnwidth]{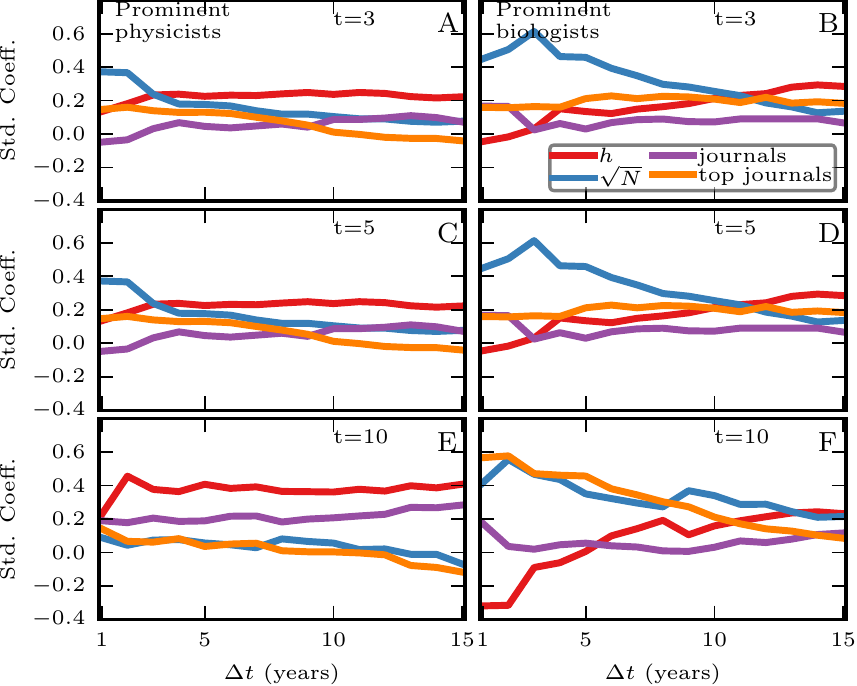}
\caption{Variation of the standard coefficient indicating the change in the contribution of each factor over $\Delta t$, the period for the calculation of $\Delta h$. 
The top panels show the coefficients for early careers ($t=3$), the mid range careers ($t=5$), and late careers ($t=10$).}
\label{fig:Rsq_dh_CI}
\end{figure}

\begin{figure}[h!]
\centering
\includegraphics[width=\columnwidth]{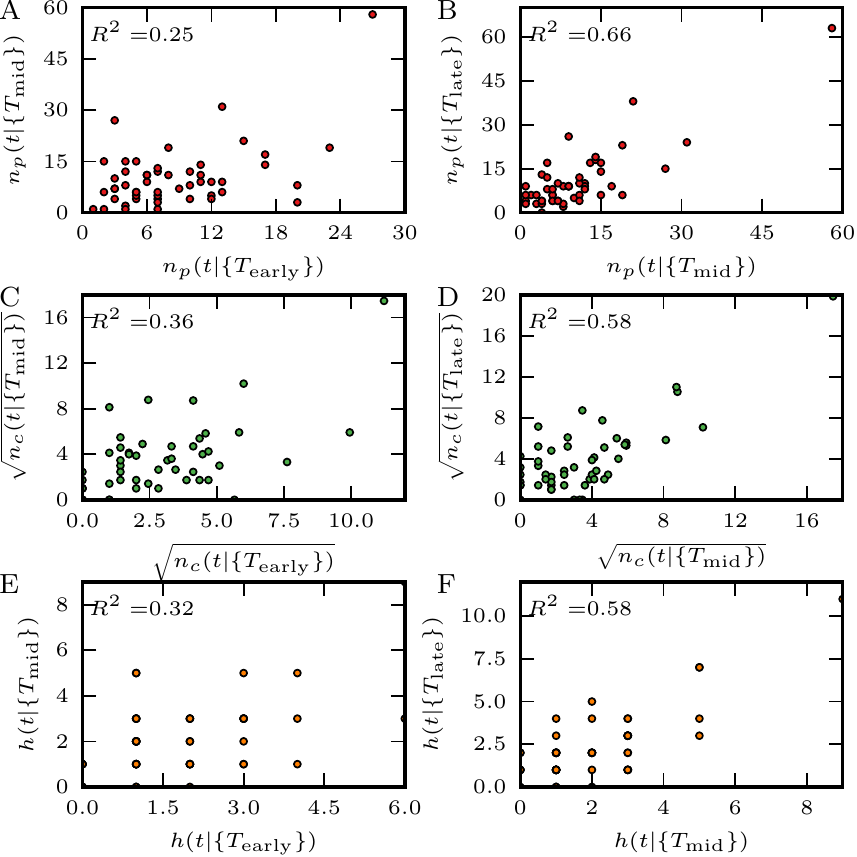}
\caption{Correlation between the past and the true future for prominent mathematics careers. Scatter plot of the (A,B) number of papers, (C,D) number of total citations, and (E,F) non-cumulative $h$-index $h(t \vert \{T_{j}\} )$ calculated for each author using non-intersecting sets of papers published in ``early''- ``mid''-  and ``late''- career periods. The correlation coefficient $R$ for each plot is also shown. The plot also confirms that for mathematicians, fluctuations in impact measure is low and hence many data points lay on top of each other.}
\label{fig:trueFutureMaths}
\end{figure}
\begin{figure}[h!]
\centering
\includegraphics[width=\columnwidth]{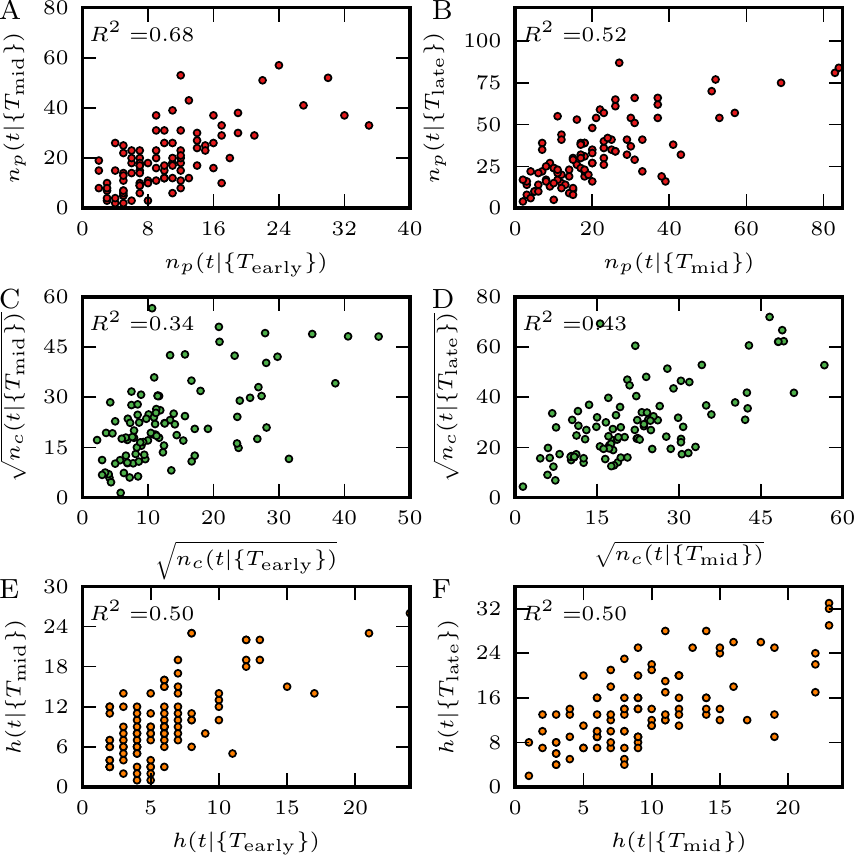}
\caption{Correlation between the past and the true future for prominent biology careers. Scatter plot of the (A,B) number of papers, (C,D) number of total citations, and (E,F) non-cumulative $h$-index $h(t \vert \{T_{j}\} )$ calculated for each author using non-intersecting sets of papers published in ``early''- ``mid''-  and ``late''- career periods.
}
\label{fig:trueFutureTopBiology}
\end{figure}

\begin{figure}[h!]
\centering
\includegraphics[width=\columnwidth]{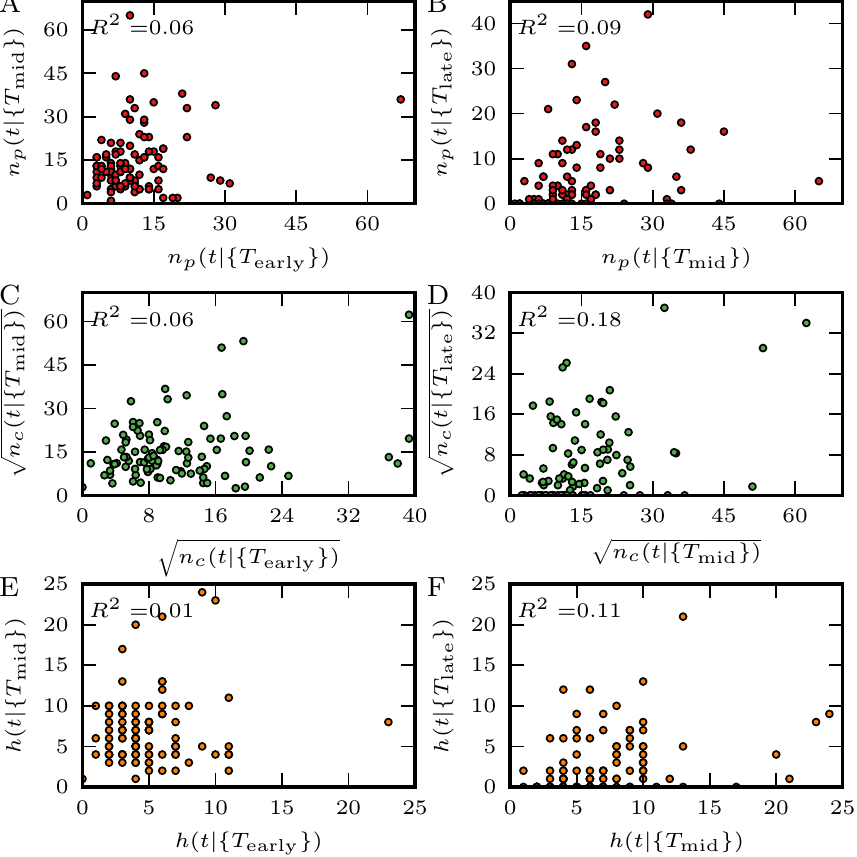}
\caption{Correlation between the past and the true future for assistant professors in physics. Scatter plot of the (A,B) number of papers, (C,D) number of total citations, and (E,F) non-cumulative $h$-index $h(t \vert \{T_{j}\} )$ calculated for each author using non-intersecting sets of papers published in ``early''- ``mid''-  and ``late''- career periods.}
\label{fig:trueFuturePhysAsst}
\end{figure}

\begin{figure}[h!]
\centering
\includegraphics[width=\columnwidth]{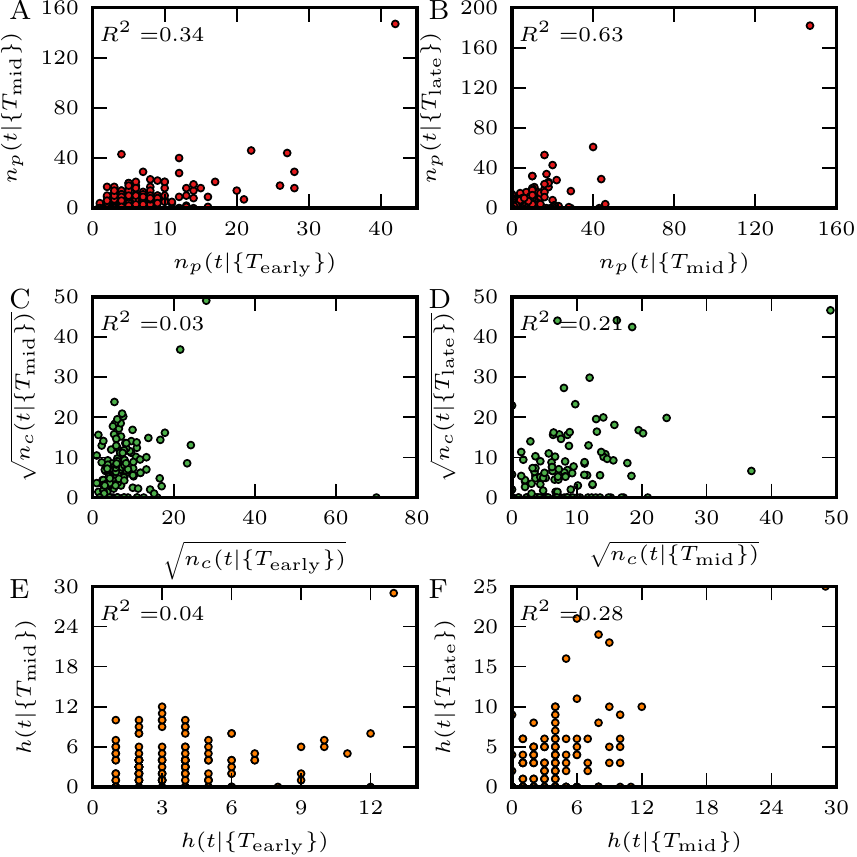}
\caption{Correlation between the past and the true future for careers in biology. Scatter plot of the (A,B) number of papers, (C,D) number of total citations, and (E,F) non-cumulative $h$-index $h(t \vert \{T_{j}\} )$ calculated for each author using non-intersecting sets of papers published in ``early''- ``mid''-  and ``late''- career periods. The correlation coefficient $R$ for each plot is also shown.}
\label{fig:trueFutureBiology}
\end{figure}
\begin{figure}[h!]
\centering
\includegraphics[width=\columnwidth]{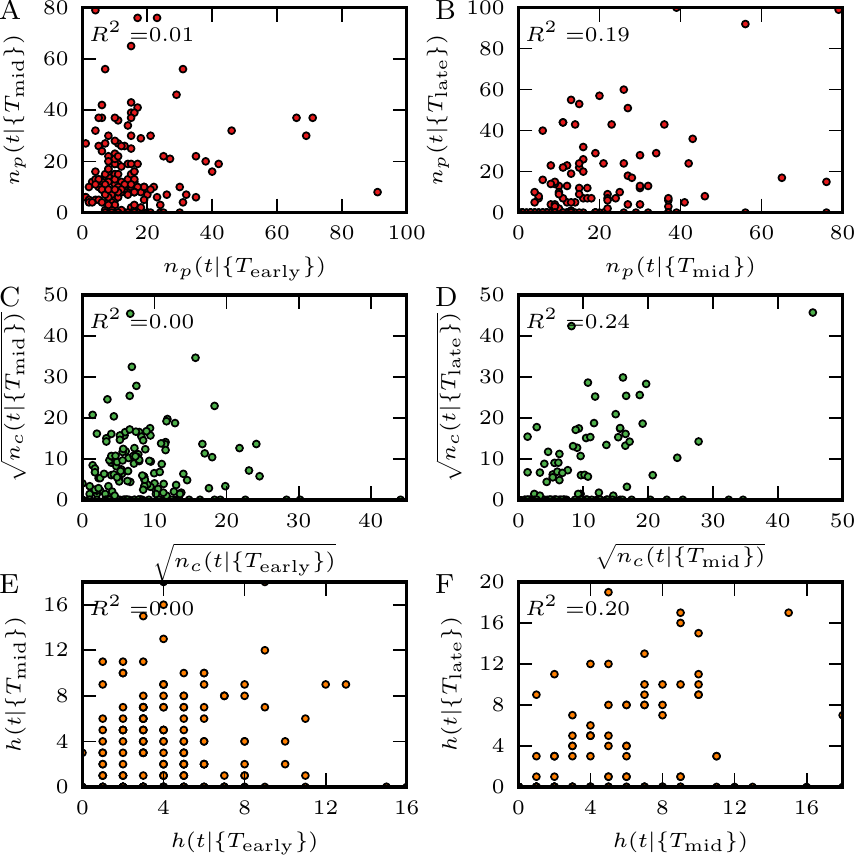}
\caption{Correlation between the past and the true future for graphene researchers. Scatter plot of the (A,B) number of papers, (C,D) number of total citations, and (E,F) non-cumulative $h$-index $h(t \vert \{T_{j}\} )$ calculated for each author using non-intersecting sets of papers published in ``early''- ``mid''-  and ``late''- career periods. The correlation coefficient $R$ for each plot is also shown.}
\label{fig:trueFutureGraphene}
\end{figure}
\clearpage
\end{document}